\documentclass[aps,prl,twocolumn,groupedaddress,showpacs,showkeys,amsmath,amssymb]{revtex4-1}

\usepackage{graphicx}
\usepackage{bm}
\newcommand{\unit}[1]{\mathbf{\hat{#1}}}
\newcommand{\dyad}[1]{\overleftrightarrow{\mathbf{#1}}}

\begin{document}

\title{Electric dipole-free interaction of visible light with silver metadimers}

\author{P. Grahn}
\author{A. Shevchenko}
\author{M. Kaivola}

\affiliation{Department of Applied Physics, Aalto University, P.O. Box 13500, FI-00076 Aalto, Finland}

\date{\today}

\begin{abstract}
In subwavelength-sized particles, light-induced multipole moments of orders higher than the electric dipole are usually negligibly small, which allows for the light-matter interaction to be accurately treated within the electric dipole approximation. In this work we show that in a specially designed meta-atom, a disc metadimer, the electric quadrupole and magnetic dipole can be the only excitable multipoles. This condition is achieved in a narrow but tunable spectral range of visible light both for individual metadimers and for a periodic array of such particles. The electromagnetic fields scattered by the metadimers fundamentally differ from those created by electric dipoles. A metamaterial composed of such metadimers will therefore exhibit unusual optical properties.
\end{abstract}

% suggested PACS numbers in braces on next line
\pacs{}

\maketitle

In natural materials, effects of light-matter interaction, such as optical refraction and reflection, are nearly completely governed by the electric dipole excitations in the atoms and molecules composing the medium. While higher-order multipoles, such as electric quadrupoles and magnetic dipoles, can give rise to various optical phenomena \cite{RaabBook}, their contribution to the overall interaction is very small. In optical metamaterials, however, light can excite also significant magnetic dipole moments, which can lead to such extraordinary optical phenomena as negative refraction and near-field focusing \cite{Shalaev2007}.
 
Recently, it was shown that one can efficiently excite both magnetic dipole and electric quadrupole moments at visible frequencies in a pair of silver bars \cite{Cho2008}. In spite of the fact that the bars had rather large dimensions, the electric dipole contribution to the light scattering was still dominating in this structure. A metamaterial, in which the meta-atomic electric dipole moments would have a negligible contribution to the light scattering compared to the higher-order multipoles would be a unique optical material that has been neither observed in nature nor studied in scientific experiments. The radiation patterns of electric quadrupoles and magnetic dipoles fundamentally differ from those of electric dipoles. Therefore, light propagation phenomena in such higher-order metamaterials must be treated in a conceptually different way than in an ordinary medium. As an example, according to the recently introduced electromagnetic boundary conditions in the presence of electric quadrupoles, the tangential components of the electric and magnetic fields can be discontinuous across the material boundaries \cite{Graham2000,Graham2001}. Furthermore, if the electric polarization $\bm{P}$ in the material could be made negligible, the interaction of light with the material would be described by Maxwell's equations in which $\bm{P}$ is absent. Such a material can exhibit reduced optical reflection and refraction at the interface with vacuum. If, in addition, the absorption could be made low, the material itself could become essentially invisible. It is also remarkable that, in contrast to electric dipoles, the fields radiated by light-induced electric quadrupoles and magnetic dipoles in the forward and backward directions oscillate out of phase. This property could be used for creation of ultrathin optical elements, such as beam splitters and phase retarders, with unusual optical characteristics.

In this Letter, we propose a concept of metamaterial that interacts with light solely via electric quadrupole and magnetic dipole excitations. These electromagnetic multipoles are of the same order, and for brevity we refer to this type of material as quadrupole metamaterial. To facilitate a practical realization of this concept, we search for a unit structure for the material, a meta-atom, that is (i) simple, such that the material can be relatively easily fabricated, (ii) much smaller than the visible-light wavelength, for the metamaterial to be treatable as homogeneous in the visible spectral range, and (iii) symmetric, to enable a polarization independent response of the material to the field at normal incidence. To reach this goal, we employ a metadimer consisting of two closely spaced metal nanoparticles. Such dimers have previously been studied for excitation of antisymmetric \cite{Grigorenko2005,Ekinci2008,Pakizeh2008,Kante2012} or dark \cite{Zhang2008,Fan2010,Bozhevolnyi2011} plasmon modes that are characterized by simultaneous oscillation of two oppositely directed electric currents. The currents in those dimers, however, were only partially opposite and did not have equal effective amplitudes. As a result, the dominating scattering by the structures still originated from electric dipoles. In order to exactly identify and evaluate the light-induced multipole moments in our meta-atoms, we use the electromagnetic multipole expansion introduced, e.g., in Ref.~\cite{Rochstuhl2011}.

\begin{figure}[htb]
   \includegraphics[width=60mm]{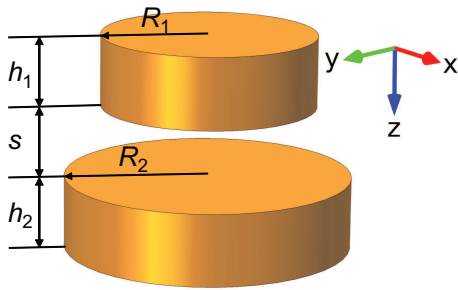}
   \caption{Illustration of the metadimer geometry.\label{figmetadimer}}
\end{figure}

The metadimer we propose as a unit of a quadrupole metamaterial consists of two axis-aligned silver discs, arranged as depicted in Fig.~\ref{figmetadimer}. The discs have the same thickness $h_1 = h_2 = 10$~nm, but different radii, $R_1 = 15$~nm and $R_2 = 20$~nm. The surface-to-surface separation of the discs is set to $s = 10$~nm. In our calculations, we choose the coordinate system to have the origin located on the dimer axis, at the center of the gap between the discs. The $z$-axis is directed such that the smaller disc is in the $z < 0$ half-space, while the larger one is in the $z > 0$ half-space. In order to have a scenario that can be realized in practice, the discs are considered to be embedded in a homogeneous dielectric of refractive index 1.5. This choice implies no loss of generality and the conclusions to be drawn are equally applicable to any other choice of lossless surrounding.

We assume that the metadimer is illuminated by a linearly x-polarized plane wave propagating in the $+\unit{z}$ direction. Because of the chosen size and geometry for the metadimer, all moments of higher order than the electric quadrupole and magnetic dipole moments can be neglected. The symmetry of the metadimer with respect to the illumination direction and polarization ensures that the electric dipole, magnetic dipole and electric quadrupole moments are $\bm{p} = \unit{x}p_x$, $\bm{m} = \unit{y}m_y$, and $\dyad{q} = (\unit{x}\unit{z}+\unit{z}\unit{x})q_{xz}$, respectively. Here $\unit x\unit z$ and $\unit z\unit x$ are the outer products of the unit vectors $\unit x$ and $\unit z$.

We first calculate the scattered electromagnetic field distribution around the metadimer using the computer software COMSOL Multiphysics. The values of the electric permittivity of silver are taken from Ref.~\cite{Johnson1972}. We then expand the scattered field by using the multipole expansion \cite{JacksonBook,Rochstuhl2011}. The expansion coefficients are used to calculate the contributions $C_s^e$, $C_s^m$ and $C_s^q$ of the electric dipole, magnetic dipole, and electric quadrupole, respectively, to the scattering cross section of the metadimer. These modal cross sections describe the amount of optical power radiated to the far-field by the corresponding multipole moment, relative to the intensity of the incident plane wave. Thus, the modal cross sections enable us to compare the different multipole excitations.

\begin{figure}[htb]
   \includegraphics[width = 85mm]{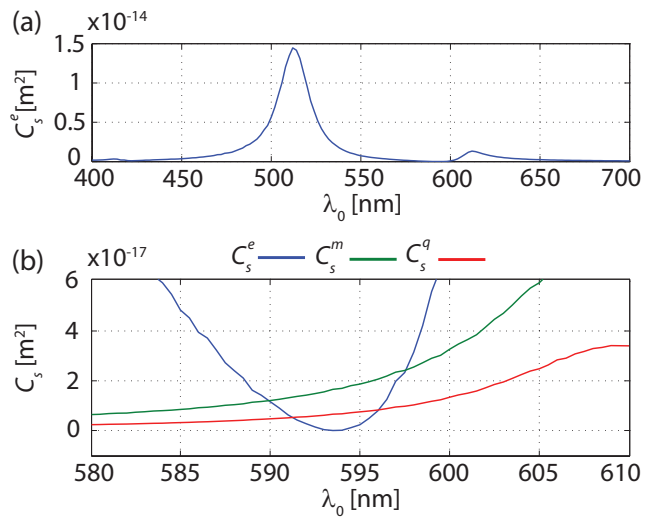}
   \caption{Spectra of the modal scattering cross sections of a silver metadimer. The metadimer is embedded in glass and has the dimensions ($R_1$,$h_1$,$R_2$,$h_2$,$s$) = (15,10,20,10,10), in nm. (a) The whole visible spectrum of the cross section $C_s^e$ due to the electric dipole mode. (b) The cross sections originating from each of the three lowest order modes around the wavelength of the electric dipole suppression. \label{figdimer}}
\end{figure}

The modal cross sections of the metadimer as a function of the vacuum wavelength $\lambda_0$ are depicted in Fig.~\ref{figdimer}. Figure~\ref{figdimer}a shows the electric dipole contribution to the scattering cross section over the whole visible spectrum. It can be seen that at around $\lambda_0 = 594$~nm, the electric dipole moment is suppressed and, as Fig.~\ref{figdimer}b indicates, the power scattered by the electric dipole mode is negligibly small compared to that scattered by the electric quadrupole and magnetic dipole modes. While the two remaining cross-sections $C_s^m$ and $C_s^q$ are small, they entirely determine the light-metadimer interaction at this wavelength.

\begin{figure}[htb]
   \includegraphics[width = 85mm]{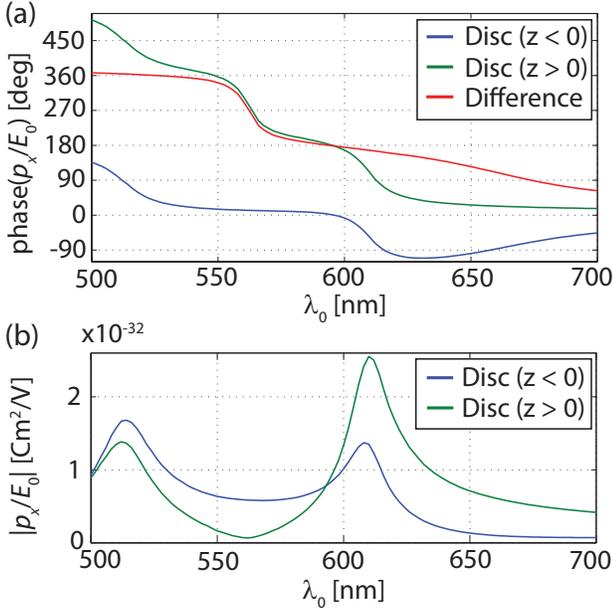}
   \caption{Spectra of (a) the phase and (b) the absolute value of the $x$-components of the excited dipole moments in the individual discs of the metadimer, normalized to the electric field amplitude $E_0$ of the incident light.\label{figdimerpx}}
\end{figure}

To obtain an intuitive picture of the electric current distribution in the metadimer, we calculate the electric dipole moments excited in each of the coupled discs. The magnitudes and phases of these moments are depicted in Fig.~\ref{figdimerpx}. In the spectral region around 594 nm, the electric dipole moments of the two discs oscillate out of phase with respect to each other (see Fig.~\ref{figdimerpx}a). A complete electric dipole suppression is obtained if, in addition, the two dipole moments have equal amplitudes. Figure~\ref{figdimerpx}b shows that the magnitudes of the dipole moments are indeed equal at the wavelength of 594 nm, which is in agreement with Fig.~\ref{figdimer}.

We have verified by calculations that increasing the separation between the two discs will blue-shift the spectral location of the electric dipole suppression. For example, at a separation of $s=30$~nm this location is shifted to a vacuum wavelength of 546 nm. Thus, varying the separation allows us to tune the spectral location of the dominating electric quadrupole and magnetic dipole scattering.

Next, we show that in the multipole expansion of the electric current in the metadimer, the electric quadrupole and the magnetic dipole are interconnected excitations of the same order. The small size of the metadimer allows us to treat it as a point particle. Neglecting all multipole moments of higher order than the electric quadrupole and magnetic dipole, the electric current density of a point particle can be written in the form (see, e.g., \cite{Russakoff1970})
\begin{equation}\label{currentJ}
\bm{J}(\bm{r}) = -i\omega\big(\bm{p} -\dyad{q}\cdot\nabla\big)\delta(\bm{r}) -\bm{m}\times\nabla\delta(\bm{r}).
\end{equation}
At $\lambda_0 = 594$~nm, the dominating excitation in the metadimer can be seen as a second order electric current excitation, a \textit{current quadrupole}. The major contribution to the excited current quadrupole can be considered to consist of two opposite x-oriented current elements of length $L$ separated by a distance $s$ along the z-axis. If the element located at $z > 0$ carries a complex-valued current of $+I_{xz}$ and the one at $z < 0$ a current of $-I_{xz}$, the electric current density in the point particle limit is \cite{HarringtonBook}
\begin{equation}\label{currentQuadrupole}
\bm{J}_{xz}(\bm{r}) = -\unit x I_{xz}Ls\frac{d}{dz}\delta(\bm{r}).
\end{equation}
Comparing Eq.~(\ref{currentJ}) with Eq.~(\ref{currentQuadrupole}), we note that in the $\bm{J}_{xz}$ excitation the non-zero multipole moments are $\dyad{q} = (\unit{x}\unit{z}+\unit{z}\unit{x})iI_{xz}Ls/(2\omega)$ and $\bm{m} = \unit{y}I_{xz}Ls/2$. Another contribution to the electric current excitation present in the metadimer is $\bm{J}_{zx}$, in which we have a similar $\dyad{q} = (\unit{x}\unit{z}+\unit{z}\unit{x})iI_{zx}Ls/(2\omega)$, but opposite $\bm{m} = -\unit{y}I_{zx}Ls/2$, where, however, $I_{zx}$ is small compared to $I_{xz}$.

\begin{figure}[htb]
   \includegraphics[width = 85mm]{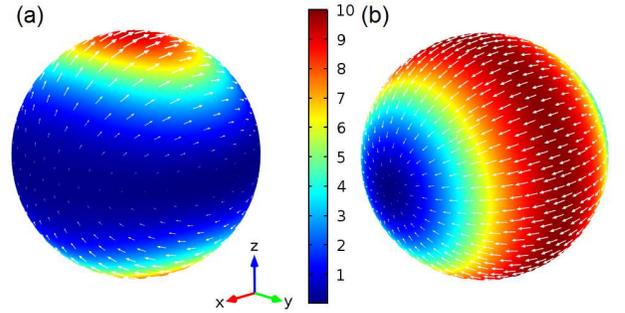}
   \caption{(a) Far-field radiation pattern of the metadimer at the wavelength of 594~nm. The excitation field is x-polarized. (b) Far-field radiation pattern of an x-oriented electric dipole. Arrows: electric field vectors. Colors: normalized intensity.\label{figfar}}
\end{figure}

It is of interest to compare the field created by the current quadrupole excitation in the metadimer to the field created by an electric dipole. In spherical coordinates, the electric far-field of a point electric dipole oriented along the x-axis and situated at the origin of the coordinate system is
\begin{equation}\label{pRad}
\bm{E}_F^{p}(\bm{r}) = \frac{k^2p_x}{4\pi\epsilon}\frac{e^{ikr}}{r}\big(\unit\theta \cos\theta\cos\phi -\unit\phi\sin\phi\big).
\end{equation}
On the other hand, the electric far-field of the point current quadrupole introduced in Eq.~(\ref{currentQuadrupole}) can be calculated to assume the form \cite{HarringtonBook}
\begin{equation}\label{qxzRad}
\bm{E}_F^{q}(\bm{r}) = \frac{I_{xz}Ls}{\omega}\frac{k^3}{4\pi\epsilon}\frac{e^{ikr}}{r}\cos\theta\big(\unit\theta \cos\theta\cos\phi-\unit\phi \sin\phi\big).
\end{equation}
In contrast to an electric dipole, which radiates symmetrically around its axis, the current quadrupole radiates primarily in the $+\unit z$ and $-\unit z$ directions. Furthermore, the fields radiated in the $+\unit z$ and $-\unit z$ directions oscillate out of phase. The numerically calculated far-field radiation pattern of the metadimer, at the wavelength of the electric dipole suppression, is depicted in Fig.~\ref{figfar}a. For comparison, Fig.~\ref{figfar}b shows the far-field radiation pattern of an electric dipole. The field profile in Fig.~\ref{figfar}a is in good agreement with Eq.~(\ref{qxzRad}).

\begin{figure}[htb]
   \includegraphics[width = 85mm]{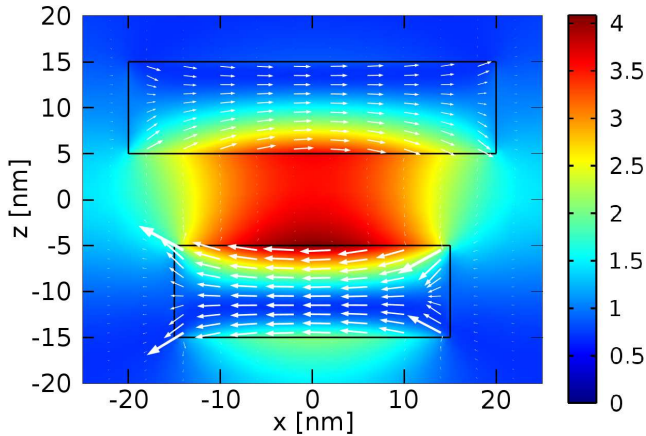}
   \caption{The electric current distribution and the magnetic field amplitude in an individual silver metadimer belonging to a two-dimensional square lattice of dimers embedded in glass with a period $\Lambda=50$~nm. The illumination wavelength is $\lambda_0 = 618$~nm. Arrows: electric current density vectors. Colors: absolute value of the complex magnetic field normalized to that of the incident wave.\label{figArray}}
\end{figure}

To create a quadrupole metamaterial that can be treated as homogeneous, the metadimers should be arranged in a lattice with a short period $\Lambda << \lambda_0$. Thus, we have studied the metadimers arranged in an infinite two-dimensional periodic lattice. The electromagnetic coupling between the adjacent metadimers was found not to hinder the realization of the electric dipole suppression, but simply to red-shift its spectral location. For example, for a square lattice with a period of 50~nm, we find the electric dipole suppression to take place at $\lambda_0 = 618$~nm. The electric current distribution (including both the conduction and polarization currents in the silver discs and in the surrounding dielectric) within a single unit cell of the lattice is depicted in Fig.~\ref{figArray}. The plot corresponds to the instant of time at which the electric currents have maximum values. The excitation in the metadimer exhibits zero electric dipole moment and is characterized by electric quadrupole and magnetic dipole moments of considerable strengths. In the figure, we also plot the amplitude of the magnetic field normalized to that of the incident wave. Within the metadimer, the magnetic field is seen to be significantly enhanced due to the strong opposite electric currents. These calculations clearly indicate the possibility to obtain a metamaterial with dominating electric quadrupole and magnetic dipole excitations.

As is evident from the symmetry of the metadimers, the response of the metamaterial layer is insensitive to the polarization of light at normal incidence. Another interesting property of the metadimer lattice is that the reversal of the illumination direction makes the dipole suppression phenomenon disappear and the lattice acts essentially as a lattice of electric dipoles. This can be used to create ultrathin bifacial optical components \cite{Lee2008}.

A three-dimensional quadrupole metamaterial can be obtained by stacking several layers of metadimers on top of each other. Owing to the geometrical simplicity of the proposed metadimers, the metamaterials composed of them should be relatively easy to fabricate. This would open up a possibility to study the properties of the still unexplored higher-order metamaterials also experimentally.

In summary, we have introduced a concept of quadrupole metamaterials in which electric quadrupoles and magnetic dipoles entirely determine the light-matter interaction phenomena. We have proposed a realistic meta-atom, in the form of a metal-disc nanodimer that can be used to construct such a metamaterial for operation at certain wavelengths of visible light. The spectral location at which the electric dipole contribution is suppressed can be changed by tuning, e.g., the disc separation. Our calculations show that dominating higher-order multipole scattering by the metadimers can also be achieved in a periodic two-dimensional array. The size of the metadimers and the period of the lattice were chosen to be much smaller than the wavelength, which allows for the metamaterial layer to be treated as optically homogeneous.

We anticipate that practical realizations of the proposed metamaterial will not only help verify theoretical predictions of light interaction with higher-order multipole materials \cite{Gunning1997,Lange2006}, but also lead to discovering new optical phenomena. Furthermore, we believe that the effects related to higher-order multipoles, such as optical activity, gyrotropic birefringence and dynamic magnetoelectric effect \cite{RaabBook}, can be substantially enhanced and controlled in quadrupole metamaterials designed specifically for these purposes.

\begin{acknowledgments}
This work was funded by the Academy of Finland (project 134029).
\end{acknowledgments}

% Create the reference section using BibTeX:
\bibliography{library}

\end{document}